\documentstyle[titlepage,fleqn,12pt]{article}
\begin{document}
\begin{titlepage}

\title{
Non-linear $I$-$V$ characteristics of double Schottky barriers
and polycrystalline semiconductors
}

\author{ E. Canessa and V.L. Nguyen
\\
\\
\\
Condensed Matter Group\\
ICTP-International Centre for Theoretical Physics\\
P.O. Box 586, 34100 Trieste\\
ITALY
}

\date{}
\maketitle

\baselineskip=17pt

\begin{abstract}

An attempt to determine theoretically
the highly non-linear current-voltage ($I$-$V$) characteristics
of polycrystalline semiconductors, such as $ZnO$-based varistors,
is made from the electrical properties of individual
grain boundaries under dc bias.
The role played by the fluctuations of double Schottky
barrier heights at grain interfaces on driving electrical
breakdown phenomena of macroscopic samples is pointed out
in terms of a binary mixture model.
An alternative trial form for the double Schottky barrier height
is introduced to reproduce
the breakdown voltage as well as the
high non-linear coefficient $\alpha$, where $I \propto V^{\alpha}$.


\end{abstract}

\end{titlepage}
\baselineskip=20pt
\parskip=0pt

A grain boundary in polycrystalline semiconductors
(PCS) becomes electrically active
as a result of charge trapping by gap states localized in
the interface between adjacent grains.
Such interface states may be created by the impurities,
dislocations or interfacial defects and give rise
to the appearance of double Schottky barriers \cite{1}.
The electrical conduction through a grain boundary Schottky barrier may be
determined by different conducting mechanisms, such as
tunneling, field emission, space-charge limit,
thermoionic emission, {\em etc.}, (see, {\em e.g.}, \cite{2}),
depending on the material and preparation.
In (commercial) $ZnO$-based varistors at room temperature,
thermoionic emission is believed to be the dominant mechanism
for electrical conduction through a
(grain-boundary-grain) GBG junction \cite{2,3,4,5}.

Nevertheless, a theoretical analysis of the
non-linear current-voltage ($I$-$V$) characteristics of a
macroscopic sample of PCS from the point of view of
electrical properties of individual GBG junctions is not trivial.
This non-linearity is due to the complex charge and current
feedback mechanisms occuring at grain boundaries
which lead to a large variety of potential barriers.
Thus, the barrier height (and, in correspondence,
the breakdown voltage) of a GBG junction differs from one to
another as shown by microelectrode measurements \cite{6,7}.
The random nature in size and position of grains
is revealed by electron micrograph experiments \cite{2,3,6,7,8}.

The simplest model proposed to characterize $ZnO$-varistors
is due to Levinson and Philipp \cite{9}.
In their model all $ZnO$-GBG junctions
are assumed to be identical and regularly arranged
in a cubic lattice.  To a crude approximation, the $I$-$V$ characteristics
of PCS then became completely determinated by the properties of one
GBG junction.  More recent studies, that included parallel or series
GBG structures or used the percolation approximation,
have also neglected fluctuations of double Schottky barrier heights
\cite{8,10}.
Thus, the question of how the non-linear coefficient $\alpha$, where
$I \propto V^{\alpha}$, and the electrical breakdown voltage of PCS
may be affected by the distribution of barrier heights at
grain boundaries is still open.

In this work we made the first attempt to determine in simple terms
the highly non-linear $I$-$V$ characteristics
of $ZnO$-based varistors under dc bias by considering
the non-linear electric characteristics of single grain boundaries.
For that purpose, we use a 3D effective medium
approximation (EMA) \cite{11,12} for a binary mixture of
grain boundary Schottky barriers, having different heights,
in conjuction with the thermoionic emission mechanism (TEM) for the
carrier transport through grain boundaries.  An alternative trial form for
the voltage dependence of a single
barrier height is also introduced.  Herein, we study as a function
of barrier probability the variations in the macroscopic
($i$) non-linear coefficient $\alpha$
and ($ii$) electric breakdown voltage $V_{B}$ of a PCS sample
featured by the binary mixture.

The quantity $\alpha$ is commonly defined as
\begin{equation}\label{eq:dd1}
\alpha (V) = \frac{d (\ln I)}{d (\ln V)} \;\;\; .
\end{equation}
For the breakdown voltage $V_{B}$, however,
there are somewhat different definitions.
For the present purposes, it is natural to define $V_{B}$ as
the voltage related to the maximum value of $\alpha (V)$,
{\em i.e.} as the solution of
\begin{equation}\label{eq:dd2}
(\frac{\partial \alpha }{\partial V})\mid_{V_{B}} = 0 \;\;\; .
\end{equation}
Accordingly, by $\alpha \equiv \alpha (V_{B})$ we shall denote the maximum
value of the non-linear coefficient.

In the traditional TEM
the current that is injected through a potential barrier is \cite{2,3,4,5,6}
\begin{equation}\label{eq:s1}
I=A^{*}T^{2}e^{-[ \; \frac{\phi (V) + E_{f} }{k_{B}T} ] }
    (1-e^{-\; \frac{eV}{k_{B}T} }) \;\;\; ,
\end{equation}
where $A^{*}$ is Richardson's constant, $V$ is voltage drop across a
junction, $E_{f} $ is the Fermi energy in the grain bulk
(counted from the bottom of the conducting band),
$T$ is the temperature, $e$ is the elementary charge
and $\phi $ is the potential barrier height that depends on $V$.

Solving the Poisson equation, within the simplest framework
of the double Schottky barrier model ({\em i.e.} neglecting
effects due to holes and interface processes), such that
the interface charge is independent of $V$ and time, $\phi (V)$
is found to be \cite{6,13}
\begin{equation}\label{eq:s2}
\phi (V) = \phi_{o}(1-\frac{eV}{4\phi_{o}} )^{2} \;\;\; ;
  \hspace{2cm} (eV \leq 4\phi_{o}) \;\;\; ,
\end{equation}
where $\phi_{o} \equiv \phi (V=0)$ is the potential barrier height at
zero applied $V$.  The voltage $V_{c}=\frac{4\phi_{o}}{e}$, that is
related to $\phi(V_{c})=0$, is often interpreted as the critical
voltage for electric breakdown in a single GBG junction \cite{6}.

Nevertheless, according to definitions (\ref{eq:dd1}) and (\ref{eq:dd2})
and to the $I$-$V$ characteristics of an individual GBG junction, {\em i.e.}
Eqs.(\ref{eq:s1}) and (\ref{eq:s2}), the breakdown voltage to a first
approximation can be estimated from
\begin{equation}\label{eq:k1}
V_{B} \approx \frac{2\phi_{o}}{e} \;\;\; ,
\end{equation}
and, in correspondence, the non-linear coefficient then becomes
\begin{equation}\label{eq:k2}
\alpha \approx \frac{\phi_{o}}{2k_{B}T} \;\;\; .
\end{equation}
Thus, the value of $V_{B}$ given in Eq.(\ref{eq:k1})
is two times less than the above $V_{c}$ value.   The
measured breakdown voltages are really about $V_{c}$ \cite{5,6,9}.
For a typical value of $\bar{\phi}_{o} \approx 0.75 \; eV$ at room
temperature, Eq.(\ref{eq:k2}) leads to $\alpha \approx 15$,
which is not high enough to explain the observed strong non-linearity
of $I$-$V$ characteristics of typical $ZnO$-based varistors
($\alpha\approx 25$ to 50 or more) \cite{9}.

In Fig.1 (left-hand side axis) the behaviour of $\phi (V)$ as given by
Eq.(\ref{eq:s2}) is compared with measured data normalized to
a single junction ($\phi_{o}=0.73 \; eV$; $T=300 \; K$) \cite{13}.
This figure shows the poor agreement obtained
with experimental results which have
only a slight voltage dependence at small values of applied $V$
and suddenly decrease in the vicinity of the breakdown region.
In this figure the principal features of Eq.(\ref{eq:s1})
using Eq.(\ref{eq:s2}) are also depicted for the same set of
parameters (right-hand side axis).  We note that at
$\frac{eV}{k_{B}T}~<~<~1$,
{\em i.e.} the prebreakdown region, the $I$-$V$ characteristics obtained
using (\ref{eq:s1}) and (\ref{eq:s2}) present an ohmic behaviour with
activation energy equal to $\phi_{o}$.  Without paying much attention
to this ohmic region, the curves in Fig.1 have been drawn to illustrate
electrical breakdown phenomena at a GBG junction.  In fact,
at $V\approx \frac{2\phi_{o}}{e}$, this figure shows (solid line,
right-hand side axis) that the current in the TEM suddenly increases
as $V$ increases in accord to Eq.(\ref{eq:k1}).
The value of $\alpha$ obtained from the curve $I(V)$ is exactly
$\frac{\phi_{o}}{2k_{B}T}$, ({\em c.f.} Eq.(\ref{eq:k2})).
For $V \geq \frac{4\phi_{o}}{e}$, it is experimentally found that
the current reaches the value limited by the
resistance of the $ZnO$ grains.  The $I$-$V$ characteristics are then
described by a second ohmic law which is not accounted by Eq.(\ref{eq:s1}).

The above discussion concerns an individual GBG junction only.
As mentioned there exists many junctions in PCS which are
not (necessarily) identical and have different $\phi_{o}$'s.  So,
we determine next by numerical approach the macroscopic
$I$-$V$ characteristics of PCS.  To achieve this we consider
a 3D-EMA for a binary mixture of GBG junctions with non-linear
$I$-$V$ characteristics and use the TEM for d.c. conduction
through single GBG junctions.

The EMA, originally proposed in Ref.\cite{11},
is a self-consistent scheme used to determine the conductivity
of inhomogeneous materials (see also \cite{12}).  It consists on
replacing the random net by a homogeneous lattice, where all lattice bonds
have the same conductance $\sum$.  Choosing the lattice spacing to be unity,
then $\sum$ is also the macroscopic conductivity of the homogeneous
material which can be derived self-consistently as discussed in
Ref.\cite{11,12}.  However,
it is important to mention here that this approach has been
initially applied to linear systems with constant bond conductances
$\sigma_{1}$, $\sigma_{2}$.  Noting that in this work $\sigma_{1}$ and
$\sigma_{2}$ of GBG junctions depend on the applied voltage $V$, it is
straighforward to repeat the EMA procedure to get a system of self-consistent
equations for determining the sample conductivity $\sum$ as a function
of $V$.  In this case,
the prediction of the EMA for $\sum$ of a 3D binary distribution of bond
conductances is obtained from
\begin{equation}\label{eq:s7}
p\; \frac{\sum -\sigma_{1}(V) }{\sigma_{1}(V^{*})+2\sum} +
    (1-p)\frac{\sum -\sigma_{2}(V) }{\sigma_{2}(V^{*})+2\sum}=0  \;\;\; ,
\end{equation}
and
\begin{equation}\label{eq:s8}
\frac{V^{*}_{i}}{V}=\frac{3\sum +\sigma_{i}(V^{*}_{i})-\sigma_{i}(V)}
    {2\sum + \sigma_{i}(V^{*}_{i})} \;\;\; \;\;\; , \;\;\; (i=1,2)
\end{equation}
where $p$ is the probability of having bonds with conductance $\sigma_{1}$.

We represent next the $ZnO$-grains as lattice sites and the $ZnO$-GBG
junctions as bonds.  Applying
Eqs.(\ref{eq:s1}) and (\ref{eq:s2}) to obtain $\sigma_{1}$ and $\sigma_{2}$,
each of which is derived assuming different values for
$\phi_{o}^{(1)}$ and $\phi_{o}^{(2)}$, we can then estimate the
conductivity of the system as a function of probability $p$
by solving self-consistenly Eqs.(\ref{eq:s7}) and (\ref{eq:s8}).

Figure 2 shows our results for the binary mixture conductivity.  These
are obtained from EMA within the probability range $0.1<p<0.9$ by
using typical data for $ZnO$-based varistors, {\em i.e.}
$\frac{\phi_{o}^{(1)}}{k_{B}T}=9$ and $\frac{\phi_{o}^{(2)}}{k_{B}T}=30$
\cite{7}.  In the event of an extremely narrow distribution of one type of
barrier heights (either p=0 or p=1), the predictions of the EMA
can be shown to be asymtotically correct \cite{12}.
To this end we emphasize again that we are dealing with a binary mixture
of grain junctions having positive $p<1$.

{}From Fig.2 we get
interesting results for the sample conductivity over a wide $p$-interval.
For $p<0.3$, the binary mixture
conductivity presents a behaviour similar to the case of single
GBG junctions with highest double Schottky barriers $\phi_{o}^{(2)}$
({\em c.f.} Fig.1).
On the other hand, when $p>0.5$, the sample $\sum$ is essentially
determined by the lowest double Schottky barriers having $\phi_{o}^{(1)}$.
In the intermediate region $0.3<p<0.7$, as discussed below,
it is worthwhile to mention that there is an assymetric contribution
to the sample electrical properties from each GBG component.

As it is well known, at the EMA percolation threshold value
for the component one, {\em i.e.}
$p_{c}^{(1)}\approx 1/3$ \cite{12}, the binary sample conductivity
slightly deviates
from a smooth increasing on varying $V$.  Similarly,
at the percolation threshold value for the second component, {\em i.e.}
$p_{c}^{(2)}\approx 1-\frac{1}{3}$, the sample $\sum$ also
deviates from a smooth behaviour on increasing $V$.  This is due
to deficiences in the EMA which is known to give unaccurate
predictions at these particular junction probabilities \cite{12}.
Furthermore it must be noted that, for all $p$-values considered,
the sample conductivity continues to be an increasing function
of applied voltage until -as we shall see next- electric breakdown
occurs.  For voltages $V>V_{B}$, $\sum$ tends to the grain conductivity
which is chosen as unit for conductivities in our calculations.

Having seen from Fig.2 that, to some extent, the EMA is able to reproduce
the typical non-linear
$I$-$V$ characteristics of $ZnO$-based varistors, it is then possible
to give an estimate of $V_{B}$ and $\alpha$ for the macroscopic samples.
In Figs.3(A) and 3(B) we summarize the $p$-dependences of $V_{B}$ and
$\alpha$ for the binary mixture, respectively.  The full lines correspond
to different (quality factor)
$\chi \equiv \frac{\phi^{(2)}_{o}-\phi^{(1)}_{o}}{\phi^{(1)}_{o}}$
of barrier heights;
namely (1) $\chi =2$, (2) $\chi =1$ and (3) $\chi =1/2$, where
$\frac{\phi^{(1)}_{o}}{k_{B}T}=10$ is fixed throughout calculations.
The three curves displayed in each figure are respectively normalized to
the electrical breakdown voltage
$\frac{eV_{B}^{(1)}}{k_{B}T}=\frac{2\phi_{o}^{(1)}}{k_{B}T}=20$
and to the non-linear coefficient
$\alpha^{(1)}=\frac{\phi_{o}^{(1)}}{2k_{B}T}=5$
due to component one.  Since we have chosen
$\phi^{(2)}_{o}>\phi^{(1)}_{o}$, from Eqs.(\ref{eq:k1}) and
(\ref{eq:k2}) we then have $V_{B}^{(2)}>V_{B}^{(1)}$ and
$\alpha^{(2)}>\alpha^{(1)}$.  At $p=0$, the sample $V_{B}$ and $\alpha$
become identical to those values derived from a junction with
barrier height $\phi_{o}^{(2)}$.  At $p\neq 0$, it is clear that
the sample $V_{B}$ and $\alpha$ varies between
$V_{B}^{(2)}\rightarrow V_{B}^{(1)}$ and
$\alpha^{(2)}\rightarrow \alpha^{(1)}$.

As visualized from Figs.3(A) and 3(B), the sample $V_{B}(p)$ and
$\alpha (p)$ exhibit sharp variations around $p=1/2$, especially
so at $\chi \geq 1$.  When $\chi <<1$,
a smooth dependence on $p$ of $V_{B}$ and $\alpha$ is found on decreasing
the probability of $\alpha^{(1)}$-component.  In the present cases
the sample $\alpha$ reaches a maximum value of
$\alpha^{(2)}\approx \frac{\phi_{o}^{(2)} }{2k_{B}T}$.
It is important to note that, when the conductances
(or, alternatively, $\phi_{o}^{(1)}$ and $\phi_{o}^{(2)}$)
of the components of the binary mixture
have nearly same probability $p\approx 0.5$,
there is an asymmetric contribution of the components to the non-linear
coefficient and electric breakdown voltage of the sample.
This asymmetry becomes large on increasing $\chi$.  These results
imply that the components play different roles
on driving macroscopic $I$-$V$ characteristics.

The component with barrier height $\phi_{o}^{(2)}$ becomes
important only at $p\leq 0.5$ and dominant for $p<p_{c}^{(1)}$.
On the other hand, for all $p \geq 0.5$, the sample $\alpha$ and $V_{B}$ are
enterely determined by the component
with lower barrier height, {\em i.e.} $\phi_{o}^{(1)}$,
up to an applied voltage $V$ such that
the second barrier conductivity can not be neglected.
Simultaneous contributions from both components at higher $V$
determine the behaviour of sample ($I$-$V$) characteristics
in the range $p\geq 0.7$.
Such assymetric features of the electrical properties of PCS
follow from the monotonic behaviour of the $I$-$V$
curves shown in Fig.2.  Moreover,
the sample $\alpha$ in Fig.3 is dominated by the potential barriers with
average or lower heights.  The higher double Schottky barriers do not play
considerable role in this case.

In the experiments by Olsson and Dunlop \cite{7},
several types of junctions between $ZnO$ grains were distinguished.
The various types of junctions characterized by different intergrain
structures revealed different $I$-$V$ behaviour.  The most frequent
interface structures were found to be of "type A", {\em i.e.}
$ZnO$ boundaries containing a thin intergranular amorphous
$B_{i}$-rich film, "type B", {\em i.e.} boundaries between $ZnO$ grains
not involving any second intergranular phase but containing
$Bi$ atoms, and "C type", {\em i.e.} $ZnO$ grains separated by an
integranular region containing $Bi_{2}O_{3}$.  Since types A and B
exhibited similar $I$-$V$ characteristics,
the present theoretical GBG junctions with $\phi_{o}^{(1)}$
may be related to the "C-type" junctions whereas the GBG junctions with
$\phi_{o}^{(2)}$ to the "B-type" junctions (by neglecting possible
differences in the bias dependence for these types of GBG).
Further motivation to approximate macroscopic PCS as a binary
mixture of grain boundary Schottky barriers is obtained from the experiments
by Tao {\em et al.} \cite{8}.  Our theoretical results in this case
may represent the large majority of grain-to-grain junctions
characterized from an electrical point of view as "good" junctions
and "bad" junctions.  As can be seen in Fig.3, our work predicts
that for large proportions of "good" GBG, {\em i.e.} no less than $70\%$,
the sample non-linear $I$-$V$ characteristics become completely determined
by such junctions.

The presence of free electrons in the conduction band, the hole current
(created by hot electrons at high enough applied voltages),
and interface processes (electron captures, emission and
hole-electron recombination at interface states)
(see {\em e.g.} \cite{3,4,14,15}), might be important to derive a more
satisfactory expression for the barrier height $\phi (V)$ than
Eq.(\ref{eq:s2}).  The new expression should lead to obtain
higher values of $V_{B}$ and $\alpha$ for GBG junctions, and from it,
for the $ZnO$-based varistors too.  This complicated problem has been solved
numerically for a single GBG junction \cite{3,15}.  Up to now, there is
not an analytical expression for $\phi (V)$ which, on the one hand, gives
desired results for a single junction problem \cite{16,17}
and, on the other hand,
could be basic for studing macroscospic non-linear $I$-$V$ characteristics
of $ZnO$-based varistors as it concerns us here.

As a simple example, let us consider the alternative trial form
\begin{equation}\label{eq:8x}
\phi (V)  = \phi_{o}  e^{\beta_{2} \frac{eV}{\phi_{o}} }
        (1-\beta_{1} \frac{eV}{\phi_{o}} )^{2}    \;\;\; ;
  \hspace{2cm} (\beta_{1}eV \leq \phi_{o}) \;\;\; .
\end{equation}
It can be seen in Fig.1 by dotted lines (left-hand side axis) that
this choice (though not unique) allows us to fit better experimental data
\cite{13} than using Eq.(\ref{eq:s2}) from a simple adjustment of two
parameters, namely $\beta_{1}$ and $\beta_{2}$.
Clearly, the first term of an expansion series of the exponential function
in Eq.(\ref{eq:8x}) leads to Eq.(\ref{eq:s2}) for low $V$.  The following
terms in such a series might represent those
additional effects that become
important at the high voltage region $\beta_{2} eV \approx \phi_{o}$.
Theoretical $I$-$V$ characteristics derived
using Eq.(\ref{eq:s1}) for the thermoionic emission
current and Eq.(\ref{eq:8x}) for $\phi (V)$
are also illustrated in Fig.1 by dotted lines (right-hand side axis)
for completness.

If we take $\beta_{1}$ and $\beta_{2}$ such that,
at $\frac{eV}{k_{B}T}<< 1$, the barrier heigh
becomes weakly dependent on $V$, ({\em i.e.}
$\frac{\partial \phi }{\partial V }\mid_{_{V\rightarrow 0}} \approx 0$).
Then, we immediately obtain $\beta_{2}=2\beta_{1}$ and Eq.(\ref{eq:8x})
for $\phi (V)$ becomes dependent only on $\beta_{1}$ which may be
chosen to correlate the voltage $V_{B}$
($\approx \frac{4\phi_{o}}{e}$), with a high non-linear coefficient
$\alpha$ for the GBG junctions.

In fact, from definitions (\ref{eq:dd1}) and (\ref{eq:dd2}),
for enough high barriers we have
\begin{eqnarray}\label{eq:ll1}
V_{B} & \approx & \frac{\phi_{o}}{\sqrt{2}\beta_{1}e} \;\;\; ,
        \nonumber \\
\alpha & \approx & 32 \beta_{1}^{2}(1-4\beta_{1})
               e^{8\beta_{1}}\frac{\phi_{o}}{k_{B}T}   \;\;\; .
\end{eqnarray}
For $\beta_{1}\approx 0.19$ we get the desired $V_{B}$
and $\alpha \approx 1.28 \frac{\phi_{o}}{k_{B}T}$
which is more than twice higher than $\alpha$ from Eq.(\ref{eq:k2}).
The coefficient $\alpha$
becomes independent of $\beta_{1}$, but depends on the zero-voltage
barrier heigh $\phi_{o}$.  If, as before, $\phi_{o}\approx 0.75 \; eV$ then
Eq.(\ref{eq:ll1}) gives $\alpha \approx 35$ for a single GBG junction
at room temperature.  Hence, we feel that Eq.(\ref{eq:8x})
may be a convenient expression to mimic $\phi (V)$ in these systems
in the absence of simpler approaches to include
effects due to holes and interface processes.

Let us focus again briefly on the EMA.  In Figs.3(A) and 3(B)
(dotted lines) we plot new values of the sample $V_{B}(p)$
and $\alpha (p)$ by using Eqs.(\ref{eq:s1}) together with
Eq.(\ref{eq:8x}) and $\chi =0.5$ for a comparison with the solid lines 3.
We immediately note that the behaviour of the sample $\alpha$ and $V_{B}$
is the same than in the previous case by applying Eq.(\ref{eq:s2}).
But, even more important, according to relations in Eq.(\ref{eq:ll1})
$\alpha$ and $V_{B}$ reach higher values.  The same conclusion
can be drawn for different values of $\chi$ (not shown).

To conclude some comments are in order.
We have analysed the non-trivial problem of
the non-linear $I$-$V$ characteristics of a
macroscopic sample of PCS from the point of view of
electrical properties of individual GBG junctions.
In real percolation systems a network of non-linear elements at breakdown is
known to be filamentary.  In order to take into account of this effect,
a more sophisticated model which includes
random fluctuations of (multiple) barrier heights must be considered.
In this work we have restricted ourselves to a simple binary mixture model
of PCS ($ZnO$)-grain interfaces and used the TEM for the
carrier transport through the grain boundaries.  The use of a binary mixture
approach being motivated by the experimental findings of Refs.\cite{7,8}.
In the case of a narrow distribution in fluctuations
of barrier heights \cite{6,20}, the $I$-$V$
characteristics of varistors may became very similar to the single junction
data and the simplest model of Levinson and Philipp suffices \cite{9}.

We have found that features in the $V$-dependence of $\phi$ and
fluctuations of double Schottky barrier heights at
(rather than the electrical conduction through) grain boundaries play an
essential role to achieving highly, non-linear $I$-$V$ characteristics
of macroscopic PCS samples.  For this reason, we have introduced
a simple trial form for the voltage dependence of a single
barrier height in the absence of simpler approaches to mimic
effects due to holes and interface processes.
We have assumed that the bias dependence of GBG is the same for
both components by considering either
Eq.(\ref{eq:s2}) or Eq.(\ref{eq:8x}).
As shown in Fig.3, different $\phi (V)$ functions will clearly
lead to obtain different results depending on the chosen form.
The 3D-EMA for non-linear $I$-$V$ characteristics
of PCS as introduced in this work presents an asymmetric contribution of
barrier heights to the sample $\alpha$ and $V_{B}$ around $p=0.5$.
This asymmetry becoming large on increasing the difference in
the barrier height values of the components.

These interesting findings may be useful guidelines
to simulation studies of electric breakdown to a network of random
potential barriers with non-linear $I$-$V$ characteristics \cite{18,19,21}.
In addition, similar EMA calculations may be also
extended to 2D systems such as polycrystalline semiconductor films
\cite{5}.  In the latter, we shall expect analogous results
because 2D-EMA equations are equivalent to those in
Eqs.(\ref{eq:s7}) and (\ref{eq:s8}) for 3D samples
but with a small change of coefficients.

\vspace{2cm}
\section*{Acknowledgements}

The authors benefited from a discussion with Prof. L. Pietronero
and Dr. B. Tanatar concerning electric breakdown phenomena.
Sincere thanks are extended to the referee for pointing out
important comments which resulted in a better presentation of these ideas.
The Condensed Matter Group and the Computer Centre at ICTP, Trieste, are
also acknowledged for support.

\newpage

\section*{Figure captions}

\begin{itemize}
\item {\bf Figure 1}:  Double-Schottky barrier height $\phi$ at, and
dc current through, a single grain boundary as a function of applied
voltage $V$ for $\phi_{o}=0.73\; eV$ at $T=300 \; K$.  Here
$I_{o}=A^{*}T^{2}e^{-\; \frac{E_{f}}{k_{B}T} }$.
Full lines are obtained from Eqs.(\ref{eq:s1}) and
(\ref{eq:s2}) and dotted lines from Eqs.(\ref{eq:s1}) and (\ref{eq:8x}).
Squares represent experimental data for $\phi (V)$ \cite{13}.

\item {\bf Figure 2}:  Results for the binary mixture conductivity $\sum$
as a function of applied voltage $V$ for several probabilities $p$
of having bonds with conductance $\sigma_{1}$.
Typical data for $ZnO$-based varistors are used:
$\frac{\phi_{o}^{(1)}}{k_{B}T}=9$ and
$\frac{\phi_{o}^{(2)}}{k_{B}T}=30$.

\item {\bf Figure 3}: (A) Normalized voltage $V_{B}$ at which electrical
breakdown occurs and (B) normalized non-linear coefficient $\alpha$
for a binary mixture as a function of probability $p$ of having component
one.  In these plots (1) $\chi =2$, (2) $\chi =1$ and (3) $\chi =1/2$.
Full lines are obtained using Eqs.(\ref{eq:s1}) and (\ref{eq:s2}).
Dotted lines are obtained using Eqs.(\ref{eq:s1}) and (\ref{eq:8x})
with $\chi =1/2$.

\end{itemize}

\end{document}